\documentclass[a4paper,12pt]{article}

\usepackage{amsmath,amssymb}
\usepackage{cite}

\begin{document}

\author{Nikolay A. Kudryashov\footnote{E-mail: kudryashov@mephi.ru}}

\title{Unnecessary Exact Solutions of Nonlinear Ordinary Differential Equations}

\date{Department of Applied Mathematics, \\ National  Research Nuclear University MEPHI,  \\ 31 Kashirskoe Shosse, 115409, Moscow, \\ Russian Federation}

\maketitle

\begin{abstract}

We analyze the paper by Wazwaz and Mehanna [Wazwaz A.M., Mehanna M.S., A variety of exact travelling wave solutions for the (2+1) -- dimensional Boiti -- Leon -- Pempinelli equation, Appl. Math. Comp. 217 (2010) 1484 -- 1490]. Using the tanh -- coth method and the Exp -- function method the authors claim that they have found exact solutions of the (2+1) -- dimensional Boiti -- Leon -- Pempinelli equation. We demonstrate that the authors have obtained the exact solutions of the well known nonlinear ordinary differential equation. We illustrate that all solutions presented by the authors can be reduced to the well-known solutions. Wazwaz and Mehanna made a number of typical mistakes in finding exact solutions of nonlinear differential equations. Taking the results of this paper we introduce the definition of unnecessary exact solutions for the nonlinear ordinary differential equations.

\end{abstract}

\section{Introduction}

The construction of exact solutions for nonlinear differential equations is an important part of nonlinear science and we can see many achievements in this area in the last years \cite{Kudr88, Kudr90, Parkes96, Fan01, Polyanin05, Kudr05}. Many of these achievements were reached using symbolical calculations with application of such software as MAPLE and MATEMATICA. However there are some shortcomings of this approach. Though computers can help investigators to do some calculating routine but they cannot completely replace the investigators since computers do not know mathematics. The total reliance on computers can lead to various mistakes in finding exact solutions of nonlinear differential equations.

We have seen many papers from different journals with such examples but for this note we selected one of them. Our aim is to demonstrate mistakes which Wazwaz and Mehanna  made in Ref. \cite{Wazwaz} in finding exact solutions for the system of nonlinear differential equations.

In the recent paper \cite{Wazwaz} Wazwaz and Mehanna considered a system of equations
\begin{equation}
u_{ty}= (u^2-u_x)_{xy}+2\,v_{xxx}
\label{eq1}
\end{equation}
\begin{equation}
v_{t}= v_{xx}+2\,u\,v_x
\label{eq1a}
\end{equation}

To look for exact solutions of system \eqref{eq1} -- \eqref{eq1a} the authors used the traveling wave solutions $u(x,y,t) =u(\xi)$, $v(x,y,t)=v(\xi)$, $\xi=\mu(x+y-c\,t)$. Wazwaz and Mehanna solved the following system of nonlinear ordinary differential equations
\begin{equation}
-c\,u^{''}=(u^2)^{''}-\mu\,u^{'''}+2\mu\,v^{'''},
\label{eq2}
\end{equation}
\begin{equation}
-c\,v^{'}=\mu\,v^{''}+2\,u\,v^{'},
\label{eq2a}
\end{equation}
where
\begin{equation*}
u^{'}=\frac{du}{d\xi},\quad v^{'}=\frac{d\,v}{d\xi}
\label{eq3}
\end{equation*}
and so on.

The authors have written \cite{Wazwaz}: "integrating the first equation twice with respect to $\xi$
gives"
\begin{equation}
v^{'}=\frac12\,u^{'}-\frac{u^2+c\,u}{2\,\mu}
\label{eq3a}
\end{equation}

In fact, after integration Eq.\eqref{eq2} twice with respect to $\xi$ we obtain
\begin{equation}
v^{'}=\frac12\,u^{'}-\frac{u^2+c\,u}{2\,\mu}+C_1\,\xi+C_2,
\label{eq3b}
\end{equation}
where $C_1$ and $C_2$ are arbitrary constants.
Wazwaz and Mehanna  omitted constants of integration in \eqref{eq3b} and
reduced a class of exact solutions for Eqs.\eqref{eq1}--\eqref{eq1a}. We see that the authors made the third error in finding exact solutions of nonlinear differential equations from the Kudryashov's list \cite{Kudryashov09c}.

Substituting \eqref{eq3a} into \eqref{eq2a} Wazwaz and Mehanna obtained the equation in the form
\begin{equation}
\mu^{2}\,u^{''}-2\,u^{3}-3\,c\,u^{2}-c^{2}\,u=0
\label{Jacobi}
\end{equation}
The authors of Ref. \cite{Wazwaz} investigated this second-order ordinary differential equation
using the tanh-coth function and the Exp-function methods for finding exact solutions.

The outline of this note is following: in Section 2 we give the general solution of \eqref{Jacobi}.
In Section 3 we analyze the application of the tanh - coth method for finding exact solutions of Eq.\eqref{Jacobi} and show that all solutions presented by Wazwaz and Mehanna can be reduced to one solution. In Section 4 we consider the application of the Exp - function method to Eq.\eqref{Jacobi} and illustrate that all solutions found with this method can be simplified. In Section 5 we introduce the definition of the unnecessary exact solution for a nonlinear ordinary differential equation and prove the theorem about scientific dishonesty.

\section{General solution of Eq.\eqref{Jacobi}}

Let us show that the general solution of Eq. \eqref{Jacobi} is expressed via Jacobi elliptic function.
Also we will show that all other exact solutions can be found from the general solution.

Multiplying Eq.\eqref{Jacobi} on $u^{'}$ and integrating Eq. \eqref{Jacobi} once with respect to $\xi$
we have
\begin{equation}
\mu^{2}\,(u^{'})^{2}=u^{4}+2\,c\,u^{3}+\,c^{2}\,u^{2}-\alpha
\label{Jacobi_1}
\end{equation}
where $\alpha$ is an integration constant.

Eq.\eqref{Jacobi_1} has the following general solution
\begin{equation}
u=\frac{\sqrt{c^{2}-4\,\sqrt{\alpha}}}{2}\,\mbox{sn}\left\{\frac{1}{2\,\mu}\,\sqrt{c^{2}+
4\,\sqrt{\alpha}}(\xi-\xi_{0}),\sqrt{\frac{c^{2}-4\,\sqrt{\alpha}}{c^{2}+4\,\sqrt{\alpha}}}\right\}-
\frac{c}{2},
\label{elliptic}
\end{equation}
where $\xi_0$ is an arbitrary constant.

In the case $\alpha=0$ from Eq. \eqref{elliptic} we have
\begin{equation}
u=-\frac{c}{1+\,C_1\,{\rm\,e}^{\pm\frac{c}{\mu}\eta}},
\label{S1}
\end{equation}
where $C_1$ is an arbitrary constant.

Note that solution \eqref{S1} can be presented in the form
\begin{equation}
u=\frac{c}{2}\left(\pm \tanh\left\{\frac{c}{2\,\mu}(\xi-\xi_{0})\right\}-1\right).
\label{S1a}
\end{equation}

In the case $\alpha=\frac{c^{4}}{16}$ solution of Eq. \eqref{elliptic} has the following form
\begin{equation}
u=-\,\frac{2\,c^{2}\,C_{2}\,e^{\mp\,\frac{\sqrt{-c^{2}}}{\sqrt{2}\mu}\eta}}
{C_2^2\,e^{\,\mp\,\frac{2\sqrt{-c^{2}}}{\sqrt{2}\mu}\,\eta}+2\,c^{2}}-\frac{c}{2}.
\label{S2}
\end{equation}

All exact solutions of Eq.\eqref{Jacobi} can be obtained from these solutions. However Wazwaz and
Mehanna have found the exact solutions of this equation using the tanh -- coth and Exp -- function methods.

\section{Analysis of application of the tanh-coth method to Eq.\eqref{Jacobi} by Wazwaz and Mehanna}

Let us show that all solutions of Eq. \eqref{Jacobi} found by Wazwaz and Mehanna with the tanh -- coth method can be obtained from the general solution \eqref{elliptic}. So, the authors of Ref. \cite{Wazwaz} have
made the second error from the Kudryashov's list of errors \cite{Kudryashov09c}.

At the beginning Wazwaz and Mehanna applied the tanh -- coth method to obtain solutions of Eq.\eqref{Jacobi}. They have found the following three solutions
\begin{equation}
u_{1}(\xi)=-\frac{c}{2} \pm \frac{c}{2}\,\tanh{\xi}
\label{W1}
\end{equation}
\begin{equation}
u_{2}(\xi)=-\frac{c}{2} \pm \frac{c}{2}\,\coth{\xi}
\label{W2}
\end{equation}
\begin{equation}
u_{3}(\xi)=-\frac{c}{2}\pm \frac{c}{4}\,\tanh{\xi}  \pm \frac{c}{4}\,\coth{\xi}
\label{W3}
\end{equation}
Any of these solutions can be obtained as a special case of solution \eqref{elliptic}. Moreover,
all these solutions can be reduced to the solution \eqref{S1a} of Eq.\eqref{Jacobi}.

Let us obtain solutions \eqref{W1}-\eqref{W3} from the solution \eqref{S1a}. In the case of
$\mu=\pm \frac{c}{2}$ and $\xi_{0}=0$ from \eqref{S1a} we have the solution \eqref{W1}.

Let us note that following relations hold
\begin{equation}
\tanh{\left(\xi-\frac{i\,\pi}{2}\right)}=\coth \xi
\label{R1}
\end{equation}
\begin{equation}
\tanh{\left(\xi-\frac{i\,\pi}{2}\right)}=\frac{1}{2}\left(\tanh \frac{\xi}{2}+\coth \frac{\xi}{2}\right)
\label{R2}
\end{equation}

The solutions \eqref{W2} \eqref{W3} can be easily found using the equalities \eqref{R1} and \eqref{R2}
from the solution \eqref{S1a} with $\xi_{0}=\frac{i\,\pi}{2}$ and $\mu=\pm\frac{c}{2}$ and $\mu=\pm\frac{c}{4}$ respectively.
Therefore the authors of Ref. \cite{Wazwaz} also made the fourth error from the Kudryashov's list of errors in finding exact solutions of nonlinear differential equations.

\section{Analysis of application of the Exp -- function method to Eq.\eqref{Jacobi} by Wazwaz and Mehanna}

Using the Exp -- function method Wazwaz and Mehanna also found 16 exact solutions of Eq.\eqref{Jacobi}.
These exact solutions will be given later after first sign of equality. After these solutions we present
our transformations of solutions by authors of Ref. \cite{Wazwaz}.

Exact solutions $u_1$ and $u_3$ do not satisfy to Eq.\eqref{Jacobi}. Actually, they are wrong. Let us demonstrate that all other solutions coincide with solutions \eqref{S1} and \eqref{S2}. It is easy to see that solutions $u_2$, $u_4$, $u_5$, $u_8$, $u_{11}$ and $u_{13}$ have 2 arbitrary constants but these
solutions are not reduced to the general solution. Solutions $u_1$, $u_6$, $u_9$, $u_{14}$ and
$u_{15}$ contain 3 arbitrary constants. This situation is not possible for a second-order ordinary differential equation. Solutions $u_3$, $u_7$, $u_{10}$, $u_{12}$ and $u_{16}$ contain 4 arbitrary constants. The authors of Ref. \cite{Wazwaz} should ask each other: how can it be possible?

We can easily see that Eq.\eqref{Jacobi} has the second order. As a consequence we can obtain only two arbitrary constants for the general solution. For special case we can have less then two arbitrary
constants. It is amusing that the authors of Ref. \cite{Wazwaz} do not know this fact. So they made the seventh error from the list of errors \cite{Kudryashov09c} as well.

Now let's illustrate that solutions by Wazwaz and Mehanna can be simplified to two solutions \eqref{S1}
and \eqref{S2}.

The solution $u_1$ is wrong but if we take ${\rm e}^{-\eta}$ in place of ${\rm e}^{\eta}$ in the last expression of denominator we can transform solution $u_1$
\begin{equation}\begin{gathered}
u_{1}(\eta)=\frac{ a_{0}-cb_{-1}\,{{\rm e}^{-\eta}}}{   -{\frac {a_{0}\, \left( a_{0}+b_{0}\,c \right) {{\rm e}^{\eta}}}{{c}^{2} b_{-1}}}+b_{0}+b_{-1}\,{{\rm e}^{-\eta}} }=\\
\\
=-\frac{c\,(a_0-b_{-1}\,c\,{\rm e}^{-\eta})}{(1+\frac{b_0\,c+a_0}{b_{-1}\,c}\,{\rm e}^{\eta})(a_0-b_{-1}\,c\,{\rm e}^{-\eta})}=-\frac{c}{1+\frac{b_0\,c+a_0}{b_{-1}\,c}\,{\rm e}^{\eta}}=-\frac{c}{1+C_1\,{\rm e}^{\eta}},
\label{W4a}
\end{gathered}\end{equation}
where $C_1=\frac{b_0\,c+a_0}{b_{-1}\,c}$ is an arbitrary constant. So, we obtained that the solution $u_1$ is essentially simplified and reduced to the solution \eqref{S1} at $\mu=1$.

The solution $u_{2}$ can be reduced to the solution \eqref{S1} as well at $\mu=\frac12$ because we have
\begin{equation}
\begin{gathered}
u_{2}(\eta)=-{\frac {b_{1}\,c{{\rm e}^{\eta}}}{b_{1}\,{{\rm e}^{\eta}}+b_{-1}\,
{{\rm e}^{-\eta}}}}=-\frac{c}{1+C_{1}\,e^{-2\,\eta}},
\label{W5a}
\end{gathered}
\end{equation}
where $C_1=\frac{b_{-1}}{b_{1}}$ is an arbitrary constant.

The solution $u_3$ by Wazwaz and Mehanna do not satisfy \eqref{Jacobi} but if we substitute $b_{-1}$  for $b_1$ we will have
\begin{equation}\begin{gathered}
u_{3}(\eta)=\frac{ -b_{1}\,c{{\rm e}^{\eta}}+a_{0}}  { b_{1}\,{{\rm e}^{\eta}}+b_{0}-
{\frac {a_{0}\, \left( a_{0}+b_{0}\,c
 \right) {{\rm e}^{-\eta}}}{{c}^{2}b_{1}}}}=\\
 \\
 =-\frac{c(a_0-b_1\,c\,{\rm e}^{\eta})}{(1+\frac{a_0+b_0}{b_1\,c}\,{\rm e}^{-\eta})(a_0-b_1\,c\,{\rm e}^{\eta})}=-\frac{c}{1+\frac{a_0+b_0}{b_1\,c}\,{\rm e}^{-\eta}}=-\frac{c}{1+C_{1}\,e^{-\,\eta}},
\label{W6a}
\end{gathered}\end{equation}
where $ C_1=\frac{a_0+b_0}{b_1\,c}$ is an arbitrary constant.
In this case we have solution \eqref{S1} at $\mu=1$ again.

Solution $u_4$ can be reduced to \eqref{S2} by transformations
\begin{equation}\begin{gathered}
u_{4}(\eta)= \frac{ -{\frac {{a_{0}}^{2}{{\rm e}^{\eta}}}{4b_{-1}\,c}+a_{0}-\frac{cb_{-1}}{2}\,{\rm e}^{-\eta} }} {\frac{a_{0}^{2}{\rm e}^{\eta}}{2 c^{2}b_{-1}}+b_{-1}\,{\rm e}^{-\eta}}
=-\frac{2\,c^2\,C_2\,{\rm e}^{\eta}}{2\,c^2+C_2^2\,{\rm e}^{2\,\eta}}-\frac{c}{2},
\label{W7a}
\end{gathered}\end{equation}
where $C_2=\frac{a_0}{b_{-1}}$ is an arbitrary constant. We see that \eqref{W7a} is equivalent to \eqref{S2}.

It is easy to see that solution $u_{5}$ is equivalent to \eqref{S1} at $\mu=\frac13$
\begin{equation}
u_{5}(\eta)=-{\frac {b_{2}\,c{{\rm e}^{2\,\eta}}}{b_{2}\,{{\rm e}^{2\,\eta}}+b_{-1}\,{{\rm e}^{-\eta}}}}=-\frac{c}{1+C_{1}\,e^{-3\,\eta}},
\label{W8a}
\end{equation}
where $C_1=\frac{b_{-1}}{b_2}$ is an arbitrary constant.

We have solution \eqref{S1} at $\mu=\frac12$ for solution $u_6$ if we take into consideration the
following transformations
\begin{equation}\begin{gathered}
u_{6}(\eta)=\frac{ -b_{2}\,c{{\rm e}^{2\,\eta}}-{\frac {cb_{2}\,b_{-1}\,{{\rm e}^{\eta}}}{b_{0}}} }  { b_{2}\,{{\rm e}^{2\,\eta}}+{
\frac {b_{2}\,b_{-1}\,{{\rm e}^{\eta}}}{b_{0}}}+b_{0}+b_{-1}\,{{\rm e}^{-\eta}}}=\\
\\
=-\frac{\frac{c\,b_2}{b_0}\,{\rm e}^{2\,\eta}(b_0+b_{-1}{\rm e}^{-\eta})}{(b_0+b_{-1}{\rm e}^{-\eta})(1+\frac{b_2}{b_0}\,{\rm e}^{2\,\eta})}=-\frac{\frac{c\,b_2}{b_0}\,{\rm e}^{2\,\eta}}{1+\frac{b_2}{b_0}\,{\rm e}^{2\,\eta}}=-\frac{c}{1+\frac{b_0}{b_2}\,e^{-2\,\eta}}=\\
\\
=-\frac{1}{1+C_1\,{\rm e}^{-2\,\eta}},
\label{W9a}
\end{gathered}\end{equation}
where $C_1=\frac{b_0}{b_2}$ is an arbitrary constant.

We obtain \eqref{S1} at $\mu=1$ using the following transformations for solution $u_7$ by Wazwaz and Mehanna
\begin{equation}\begin{gathered}
u_{7}(\eta)=\frac{ -b_{2}\,c{{\rm e}^{2\,\eta}}+a_{1}\,{{\rm e}^{\eta}}+a_{0}}
{ b_{2}\,{{\rm e}^{2\,\eta}}+b_{1}\,{{\rm e}^{\eta}}-{\frac {a_{1}\,b_{1}\,c+a_{0}\,b_{2}\,c+{a_{1}}^{2}}{b_{2}\,{c}^{2}}}-{\frac
{a_{0}\, \left( a_{1}+b_{1}\,c \right) {{\rm e}^{-\eta}}}{b_{2}\,{c}^{2}}} }=\\
\\
=\frac{-b_2\,c^2\,(b_2\,c\,e^{\eta}-a_1-a_0\,e^{-\eta})\,e^{\eta}}{(b_1\,c+a_1+
b_2\,c\,e^{\eta})\,(b_2\,c\,e^{\eta}-a_1-
\,a_0\,e^{-\eta})}=\\
\\
=-\frac{c}{1+\frac{b_1\,c+a_1}{b_2\,c}\,e^{-{\eta}}}=-\frac{c}{1+C_1\,e^{-{\eta}}}.
\label{W10a},
\end{gathered}\end{equation}
where $C_1=\frac{b_1\,c+a_1}{b_2\,c}$ is an arbitrary constant.

We have solution \eqref{S1} at $\mu=\frac13$ if we simplify  $u_8$ by the authors of Ref. \cite{Wazwaz}
\begin{equation}
u_{8}(\eta)=-\frac {cb_{-1}\,{{\rm e}^{-\eta}}}{b_{2}\,{{\rm e}^{2\,\eta}}+b_{-1}\,{{\rm e}^{-\eta}}}=-\frac{c}{1+\frac{b_2}{b_{-1}}\,{\rm e}^{3\,\eta}}=-\frac{c}{1+C_1\,{\rm e}^{3\,\eta}},
\label{W11a}
\end{equation}
where $C_1=\frac{b_2}{b_{-1}}$ is an arbitrary constant.

The solution $u_9$ is reduced to solution \eqref{S1} at $\mu=\frac12$ using transformations
\begin{equation}\begin{gathered}
u_{9}(\eta)= \frac{ -b_{0}\,c-cb_{-1}\,{{\rm e}^{-\eta}}}{b_{2}\,{{\rm e}^{2\,\eta}}+{\frac {b_{2}\,b_{-1}\,{{\rm e}^{\eta}}}{b_{0}}}+b_{0}+b_{-1}\,{{\rm e}^{-\eta}} }=-\frac{c(b_0+b_{-1}\,{\rm e}^{-\eta})}{\left(1+\frac{b_2}{b_0}\,{\rm e}^{2\eta}\right)(b_0+b_{-1}\,{\rm e}^{-\eta})}=\\
\\
=-\frac{c}{1+\frac{b_2}{b_0}\,{\rm e}^{2\eta}}=-\frac{c}{1+C_{1}\,{\rm e}^{2\,\eta}},
\label{W12a}
\end{gathered}\end{equation}
where $C_1=\frac{b_2}{b_0}$ is an arbitrary constant.

Solution $u_{10}$ is reduced to the solution \eqref{S1} at $\mu=1$ taking the transformations into account
\begin{equation}\begin{gathered}
u_{10}(\eta)=\frac{ a_{1}\,{{\rm e}^{\eta}}-{\frac {c \left( cb_{2}\,b_{-1}+b_{0}\,a_{1} \right) }{a_{1}}}-cb_{-1}\,{{\rm e}^{-\eta}}} {b_{2}\,{{\rm e}^{2\,\eta}}-{\frac { \left( {c}^{3}
 {b_{2}}^{2}b_{-1}+a_{1}\,b_{0}\,b_{2}\,{c}^{2}+{a_{1}}
^{3} \right) {{\rm e}^{\eta}}}{c{a_{1}}^{2}}}+b_{0}+b_{-1}\,{
{\rm e}^{-\eta}}}=\\
\\
=\frac{c\left(a_1^3\,e^{\eta}-c\,a_1\,(c\,b_2\,b_{-1}+b_0\,a_1)-c\,a_1^2\,b_{-1}\,
e^{-\eta}\right)}{\left(\frac{b_2c}{a_1}e^{\eta}-1\right)(a_1^3\,e^{\eta}-c\,a_1\,
(c\,b_2\,b_{-1}+b_0\,a_1)-c\,a_1^2\,b_{-1}\,
e^{-\eta})}=\\
\\
=-\frac{c}{1-\frac{b_2\,c}{a_1}\,e^{\eta}}=-\frac{c}{1+C_{1}\,e^{\eta}},
\label{W13a}
\end{gathered}\end{equation}
where $C_1=-\frac{b_2\,c}{a_1}$ is an arbitrary constant.

It is easy to see that the solution $u_{11}$ is equivalent to the solution \eqref{S1} at $\mu=1$
\begin{equation}\begin{gathered}
u_{11}(\eta)=-\frac {cb_{-1}\,{{\rm e}^{-\eta}}}{b_{-1}\,{{\rm e}^{-\eta}}+b_{-2}\,{{\rm e}^{-2\,\eta}}},
\label{W14a}=-\frac{c}{1+\frac{b_{-2}}{b_{-1}}e^{-\eta}}=-\frac{c}{1+C_{1}{\rm e}^{-\eta}},
\end{gathered}\end{equation}
where $C_1=\frac{b_{-2}}{b_{-1}}$ is an arbitrary constant.

We obtain solution \eqref{S1} at $\mu=\frac12$ taking into account the transformations for $u_{12}$  in
the form
\begin{equation}
\begin{gathered}
u_{12}(\eta)=\frac{ -b_{0}\,c+a_{-1}\,{{\rm e}^{-\eta}}}{ b_{0}+b_{-1}\,{{\rm e}^{-\eta}}-
{\frac {a_{-1}\, \left( b_{-1}\,c+a_{-1} \right) {\rm e}^{-2\eta}}{c^{2}b_{0}}}}=\\=-\frac{c(b_{0}+\frac{a_{-1}}{c}\,\rm e^{-z})}{(b_{0}+
\frac{a_{-1}}{c}\,{\rm e}^{-\eta})(1+\frac{c\,b_{-1}+a_{-1}}{c\,b_{0}}\,{\rm e}^{-\eta})}=-\frac{c}{1+\frac{c\,b_{-1}+a_{-1}}{c\,b_{0}}\,{\rm e}^{-\eta}}
=\\
\\
=-\frac{c}{1+C_{1}\,{\rm e}^{-\eta}},
\label{W15a}
\end{gathered}
\end{equation}
where $C_1=\frac{c\,b_{-1}+a_{-1}}{c\,b_{0}}$ is an arbitrary constant.

We have the solution \eqref{S1} at $\mu=\frac12$ again from $u_{13}$ by the authors of Ref.\cite{Wazwaz}
\begin{equation}\begin{gathered}
u_{13}(\eta)=\frac{-c\,b_{-2}\,{\rm e}^{-2\,\eta}}{b_{0}+b_{-2}\,{\rm e}^{-2\,\eta}}
=-\frac{c}{1+\frac{b_0}{b_{-2}}{\rm e}^{\,2\,\eta}}=-\frac{c}{1+C_{1}\,{\rm e}^{\,2\,\eta}},
\label{W16a}
\end{gathered}\end{equation}
where $C_1=\frac{b_0}{b_{-2}}$ is an arbitrary constant.

We obtain the solution \eqref{S1} at $\mu=1$ using the following transformations for the solution $u_{14}$ by Wazwaz and Mehanna
\begin{equation}\begin{gathered}
u_{14}(\eta)= \frac{a_{-1}\,{\rm e}^{-\eta}-cb_{-2}\,{\rm e}^{-2\,\eta}}{ -\frac {a_{-1}\, \left( b_{-1}\,c+a_{-1} \right) }{c^{2}b_{-2}}+b_{-1}\,{\rm e}^{-\eta}+b_{-2}\,{\rm e}^{-2\,\eta}},
\label{W17a}=\\
\\
=-\frac{c\,{\rm e}^{-\eta}\left(\frac{a_{-1}}{c}-b_{-2}\,{\rm e}^{-\eta}\right)}{\left(\frac{b_{-1}}{b_{-2}}+\frac{a_1}{c\,b_{-2}}+{\rm e}^{-\eta}\right)\left(\frac{a_{-1}}{c}-b_{-2}\,{\rm e}^{-\eta}\right)}=-\frac{c\,{\rm e}^{-\eta}}{\left(\frac{b_{-1}}{b_{-2}}+\frac{a_1}{c\,b_{-2}}+{\rm e}^{-\eta}\right)}=\\
\\
=-\frac{c\,}{1+\left(\frac{b_{-1}}{b_{-2}}+\frac{a_1}{c\,b_{-2}}\,\right){\rm e}^{\eta}}=-\frac{c}{1+C_{1}\,{\rm e}^{\eta}},
\end{gathered}\end{equation}
where $C_1=\left(\frac{b_{-1}}{b_{-2}}+\frac{a_1}{c\,b_{-2}}\,\right)$ is an arbitrary constant.

We have the solution \eqref{S1} at $\mu=\frac12$ taking into account the following transformations for $u_{15}$
\begin{equation}\begin{gathered}
u_{15}(\eta)=\frac{ -c\,b_{-1}\,{\rm e}^{-\eta}-cb_{-2}\,{\rm e}^{-2\,\eta}}{ b_{1}\,{{\rm e}^{\eta}}+
\frac {b_{1}}{\,b_{-2}b_{-1}}+b_{-1}{\rm e}^{-\eta}+b_{-2}{\rm e}^{-2\,\eta}}=\\
\\
=-\frac{c\,{\rm e}^{-\eta}\,\left(b_{-1}+b_{-2}{\rm\,e}^{-\eta}\right)}{\left(\frac{b_1}{b_{-1}}{\rm\,e}^{\eta}+{\rm e}^{-\eta}\right)\left(b_{-1}+b_{-2}{\rm e}^{-\eta}\right)}=-\frac{c\,e^{-\eta}}{\frac{b_1}{b_{-1}}{\rm e}^{\eta}+{\rm e}^{-\eta}}=\\
\\
=-\frac{c}{1+C_{1}\,{\rm e}^{2\,\eta}},
\label{W18a}
\end{gathered}\end{equation}
where $C_1=\frac{b_1}{b_{-1}}$ is an arbitrary constant.

We can get the solution \eqref{S1} at $\mu=1$ from the solution $u_{16}$ by Wazwaz and Mehanna using the following transformations
\begin{equation}\begin{gathered}
u_{16}(\eta)=\frac{ a_{0}-{\frac {c \left( a_{0}\,b_{-1}+c b_{1}\,b_{-2} \right) {\rm e}^{-\eta}}{a_{0}}}-c\,b_{-2}\,{\rm e}^{-2\,\eta}}{b_{1} {\rm e}^{\eta}-\frac {a_{0}\,b_{1}\,
c^{2}b_{-1}+b_{1}^{2}c^{3}b_{-2}+a_{0}^{3}}{c\, a_{0}^{2}}+b_{-1}\,{\rm e}^{-\eta}+b_{-2}{\rm  e}^{-2\,\eta}}=\\
\\
=\frac{c\,\left(a_0^3-c\,a_0\,(a_0\,b_{-1}+c\,b_1b_{-2})\,{\rm e}^{-\eta}-c\,a_0^2\,b_{-2}
{\rm e}^{-2\,\eta}\right)}{\left(\frac{b_1c}{a_0}\,{\rm e}^{\eta}-1\right)\,\left(a_0^3-
c\,a_0\,(a_0\,b_{-1}+c\,b_1\,b_{-2}){\rm e}^{-\eta}-c\,a_0^2\,b_{-2}
{\rm e}^{-2\,\eta}\right)}=\\
\\
=\frac{c}{\frac{b_1\,c}{a_0}\,{\rm e}^{\,\eta}-1}=-\frac{c}{1+C_{1}\,{\rm e}^{\,\eta}},
\label{W19a}
\end{gathered}\end{equation}
where $C_1=-\frac{b_1\,c}{a_0}$ is an arbitrary constant.

So, we can see that all 15 solutions by Wazwaz and Mehanna from 16 ones can be simplified and reduced to the solution \eqref{S1} at $\mu=1; \,\,\,\frac12;\,\,\,\frac13$. So, the authors  \cite{Wazwaz} also made the fifth error from the list of errors from Ref. \cite{Kudryashov09c}. Now we would like to ask the authors: why did they find only 16 solutions at these values $\mu$? It is obvious that using other values $\mu$ they can find 100, 1000 or even more solutions. They can fill out all pages of the journal by these "solutions" of Eq.\eqref{Jacobi} because we have pointed out many values of parameter $\mu$. There is also another good question. Why did the authors use only multipliers with 2 monomials in the solutions $u_1$, $u_3$, $u_6$, $u_9$, $u_{12}$, $u_{14}$, $u_{15}$, with 3 monomials in the solution $u_7$, with 4 monomials in the solutions $u_{10}$ and $u_{16}$? They could obtain much more solutions for Eq.\eqref{Jacobi} if they used $5$, $6$ and more monomials. Many "new solutions" could be made up and presented as the solutions of Eq.\eqref{Jacobi}!

\section{Unnecessary exact solutions of nonlinear differential equations}

Let us introduce the definition of the unnecessary exact solution of a nonlinear differential equation.
We can observe the different types of these solutions.  Suppose we have a nonlinear differential equation
\begin{equation}
E(w, w_z, \ldots , z)=0.
\label{Eq}
\end{equation}

Assume that there is an exact solution of this equation in the form of a fraction
\begin{equation}
w=\frac{\varphi(z)}{\psi(z)}.
\label{SS1}
\end{equation}

Let us also suppose that the exact solution \eqref{S1} can be written in the form
\begin{equation}
w=\frac{\varphi_1(z)\,f(z)}{\psi_1(z)\,f(z)}=\frac{\varphi_1(z)}{\psi_1(z)}.
\label{SS2}
\end{equation}

We see that exact solution \eqref{SS1} can be simplified. We have that exact solution \eqref{SS1} is unnecessary exact solution because we can transform this solution to more simple form. Let us introduce the following definition of an unnecessary exact solution.

\emph{\textbf{Definition.} The exact solution \eqref{SS1} is called the unnecessary exact solution of the differential equation \eqref{Eq} if we can reduce this exact solution to more simple form.}

As we can see from the previous section, all the solutions of Eq.\eqref{Jacobi} obtained by Wazwaz and Mehanna using the Exp -- function method are unnecessary exact solutions. It is clear that we can obtain many solutions of nonlinear differential equation if we multiple nominator and denominator on any expression but these solutions can be simplified. All these solutions are not interesting and inconvenient. Moreover, the authors \cite{Wazwaz} do not take care of paper usage and set a bad example for many young people.

The authors of Ref. \cite{Kudryashov09a} have shown that the application of the Exp-function method in finding exact solutions is not effective, dangerous and can lead to wrong solutions. The work of Wazwaz and Mehanna illustrates all disadvantages of the Exp-function method. First, Wazwaz and Mehanna obtained wrong solutions. Second, the exact solutions from Ref. \cite{Wazwaz} contains superfluous arbitrary constants. Third, these exact solutions have cumbersome form and as we have shown above these solutions can be simplified. And the last, all correct exact solutions by Wazwaz and Mehanna are the same and can be obtained from the well-known exact solution of the Eq.\eqref{Jacobi}. So we must say to the scientific community again: be careful with Exp --   function method.

In conclusion, we strongly recommend authors to look carefully at the papers \cite{Allen, Kudryashov09c, Kudryashov09a, Kudryashov09b,  Kudryashov09d, Kudryashov09e, Kudryashov10a, Kudryashov10b, Kudryashov10c, Kudryashov10d, Kudryashov10aa, Kudryashov10bb, Kudryashov10e, Parkes09a, Parkes09b, Parkes09c, Parkes10a, Parkes10b, Popovych, Disine}  before starting to look for exact solutions of nonlinear differential equations.

\end{document}